# Online Model Estimation for Predictive Thermal Control of Buildings

Peter Radecki, *Member, IEEE,* and Brandon Hencey, *Member, IEEE*

*Abstract*—This study proposes a general, scalable method to learn control-oriented thermal models of buildings that could enable wide-scale deployment of cost-effective predictive controls. An Unscented Kalman Filter augmented for parameter and disturbance estimation is shown to accurately learn and predict a building's thermal response. Recent studies of heating, ventilating, and air conditioning (HVAC) systems have shown significant energy savings with advanced model predictive control (MPC). A scalable cost-effective method to readily acquire accurate, robust models of individual buildings' unique thermal envelopes has historically been elusive and hindered the widespread deployment of prediction-based control systems. Continuous commissioning and lifetime performance of these thermal models requires deployment of on-line data-driven system identification and parameter estimation routines. We propose a novel gray-box approach using an Unscented Kalman Filter based on a multi-zone thermal network and validate it with EnergyPlus simulation data. The filter quickly learns parameters of a thermal network during periods of known or constrained loads and then characterizes unknown loads in order to provide accurate 24+ hour energy predictions. This study extends our initial investigation by formalizing parameter and disturbance estimation routines and demonstrating results across a year-long study.

## I. INTRODUCTION

### A. Overview

Significant energy savings in buildings' heating, ventilating, and air-conditioning (HVAC) systems could be realized with advanced control systems [1], but deployment of these control systems requires a method to readily acquire low cost models of buildings' unique thermal envelopes [2], [3]. Previous studies have investigated several methods but generally arrived at non-scalable specialized solutions [4], [5], [6], [7].

Ideally, a Building Automation System (BAS) would automatically modify set-points and load shedding based on weather, occupancy, and utility pricing predictions [8]. Every building has unique and time-varying thermal dynamics, occupancy, and heat loads which must be characterized accurately if a BAS is to apply model predictive controllers (MPC) to realize energy and monetary savings [3]. Additional considerations include: measured building data often contains low information content; engineering models contain designer's intent instead of actual construction; and building's usage evolves over time [9]. Unfortunately, in practice there has yet to be demonstrated a scalable, low-cost method to readily acquire these much needed accurate models of individual buildings' unique thermal envelopes.

For continuous commissioning and lifetime adaptability a low-cost scalable method to acquire control-oriented building models must: learn both the dynamics and the disturbance patterns quickly, provide stable extrapolation, be adaptable to future changes in building structure or use, and use existing available data. White-box, first-principles, forward modeling approaches are often inaccurate, not robust to changes, and take extensive engineering or research effort to build [2], [10]. Black-box approaches take up to 6-months to train and cannot be safely extrapolated [5], [6]. Recently gray-box methods have begun to show potential as a scalable option for learning control-oriented building models in some limited specific studies [11], [12], [13], [14].

We propose a multi-mode Unscented Kalman Filter (UKF) as a generalizable on-line gray-box data-driven method to learn the building's multi-zone thermal dynamics and detect unknown time varying thermal loads. By coupling known building information and simple physics models with existing measurable building data we demonstrate how a probabilistic estimation framework can overcome shortcomings of many previously attempted specialized solutions. Our method adapts over time to continually learn both dynamics and disturbances while providing stable prediction performance.

Continuing our work in [11], this paper aims to generalize our findings and method with the following contributions:
- literature survey on control-oriented thermal modeling for buildings,
- development of minimal parameterization for dynamics estimation,
- generalized thermal disturbance pattern estimation,
- multi-mode heuristic for simultaneous parameter and disturbance estimation.

This work was supported by the Department of Defense (DoD) through the National Defense Science & Engineering Graduate Fellowship (NDSEG) Program.

P. Radecki and B. Hencey were with the Sibley School of Mechanical and Aerospace Engineering, Cornell University, Ithaca, NY, 14850 USA. Currently Radecki is a Transmission Development Engineer at General Motors Powertrain in Detroit, Michigan, *email:* ppr27@cornell.edu. Currently Hencey is with the Air Force Research Laboratory in Dayton, Ohio, *email:* bhencey@gmail.com.

A comparison of UKF and EKF applied estimation techniques is included for the benefit of practicing engineers. The robustness of the UKF estimation technique learning both parameters and disturbances is demonstrated against a multi-zone high-fidelity EnergyPlus simulation in a year-long study.

A short explanation of our proposed method follows. Using a simple first order heat transfer model with multiple zones, the UKF estimates model parameters of the thermal dynamics during periods which have small or well-characterized thermal loads. After learning the dynamics during low disturbance periods, such as nighttime, the UKF is augmented to track unknown disturbances while continuing to improve its dynamics model. The UKF is simpler than an adaptive control technique to implement because it internally maintains a covariance quality metric which only adjusts parameter estimates if the incoming data provides new thermal information. Fig. 1 shows how the proposed UKF enables rapid deployment of advanced predictive controllers for BAS.

The paper is organized as follows. First, a background section initially examines the scope of the problem and previous approaches before proposing and analyzing an extensible on-line data-driven approach. After deriving the thermal model and parameterization, we formulate and compare performance of the EKF and UKF. True utility of the UKF is then demonstrated across a year-long study. Based on data generated from our simple passive 5-zone thermal model plus a more complex passive 5-zone EnergyPlus simulated model, less than 2 weeks of training data is shown to make reliable 24-hour predictions. Based on testing and performance, a discussion of how the UKF fits into the building thermal modeling problem and an identification of areas of future research conclude the paper.

### B. Building Automation Systems Background

Buildings use 39% of the total US energy supply, a significant fraction of which is used to provide people with a comfortable indoor working and living environment by operating HVAC systems. It is estimated that 25% to 30% of building energy usage or around 10% of the total US energy consumption could potentially be reduced with component and controls upgrades [1], [2]. Reducing this energy consumption would also help significantly reduce $CO_2$ generation [9].

BAS and their connected components advanced tremendously over the past few decades and have started to be widely deployed in modern buildings and retrofit into remodeled buildings. Utilizing technologies such as wireless sensing, LonTalk, and BACnet, BAS provide networked infrastructure making it easier for sensor information and control signals to be distributed throughout and between buildings. Component advancements include sensors that can detect occupants, $CO_2$ level and light in addition to traditional temperature and humidity measurements. Furnaces, water heaters and air conditioners have all started to approach their maximum theoretical efficiency. The computational power necessary to run demand-responsive and predictive control algorithms is cheaply available.

Despite these significant advances, most modern buildings have realized only *minimal* energy savings [3]. BAS model-based predictive controllers are rarely implemented and typically underperform [15]. Larger buildings' successes in realizing energy savings has been generally limited to automated lighting control, changing nighttime temperature set-points, and load shedding during peak demand [15]. The data available from BAS is underutilized—generating trend plots instead of enabling intelligent energy management decision and control tasks [4]. It is not uncommon for buildings to simultaneously run cooling and heating components throughout all 12 months of the year [9]. Load shedding, reducing peak usage by duty cycling off portions of a system, is often based on some arbitrary component order and can significantly impact occupant comfort—if one portion of the system that is already at capacity sheds, it may not recover until nighttime [16]. Certain sites, such as Drury University and UC Merced have a human operator regularly check weather forecasts and vary temperature set points based on personal experience and intuition [16], [17]. These techniques are labor and expertise intensive. Usually, however, the cost of the human operator doesn't justify the energy savings and limits widespread deployment.

Optimal building thermal control is a multi-objective optimization problem involving user comfort, air quality, energy cost, smart grid demand response, and thermal dynamics, which generally cannot be performed optimally by a human operator without expertise. Realistic widespread improvement in building controls requires a scalable method to accurately learn unique thermal models.

### C. Building Controls and Modeling Survey

Many researchers have evaluated and implemented custom one-off BAS demonstrating energy savings with MPC [18], but widespread implementation has been illusive because a scalable method to learn thermal models has not been available. Both forward and inverse modeling efforts have traditionally taken months of a researcher's or engineer's time to accurately generate. Given that the majority of buildings which will be in use by 2030 are currently over 10 years old, the retrofit problem is significant. A scalable method must handle new construction, existing structures, and remodeling projects and continue to function if a building is repurposed.

White-box, or forward models can be generated by analyzing blueprints, building materials, and expected use patterns to generate a sophisticated simulation, but are generally not cost effective for retrofit or existing buildings. Furthermore, some studies have shown that predicted and actual energy consumption can differ by up to 40% using forward models from the design stage [10].

MPC can perform adequately given simple, accurate first or second order heat transfer models, so there is viability for inverse models [9], which use building data recorded over time to infer the dynamics. Today's networked sensors and available computation power make data-driven models feasible. The inverse modeling paradigm can use either pure numerical methods in a black-box environment or physics, first-principles based methods in a gray-box environment.

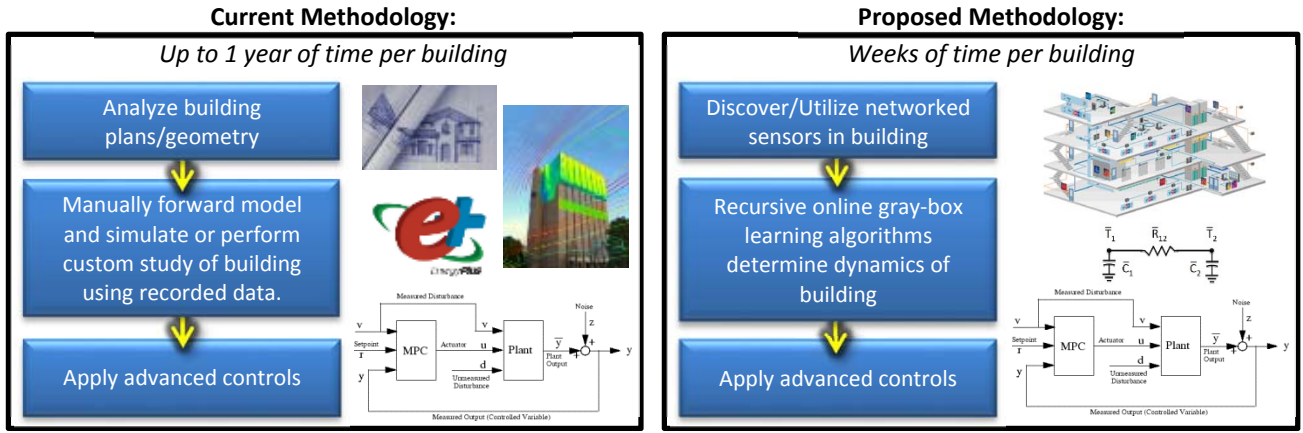

Fig. 1. Comparison of existing typical thermal modeling process *(left)* and proposed method *(right)*. Picture Credits: EnergyPlus: DOE, R. L. Smith Building (center): ThermoAnalytics, and floor plan: [3].

Many researchers from present day to those who participated in the 1990's ASHRAE Energy Predictor Shootouts have taken data-driven approaches, but their generated methods and models have typically been too specialized to scale to other buildings [2], [4], [5], [6]. A brief comparison of previously tried data-driven methods is provided before proposing our solution.

Black-box models such as Artificial Neural Networks (ANN) are often impractical for modeling thermal dynamics in building systems because of the large amount of training data (6 months to a year) that must be analyzed before getting an accurate model [6]. Because ANN create an arbitrary representation of the system, they are sensitive to the quality of the data collected, may meld together effects from loads and dynamics, are not robust to component failures, and are not adaptable when building use and configuration patterns change [2], [18]. Pure numerical methods may have better applications where simple heuristic and physical models are impractical such as pattern recognition of occupancy, lighting, or thermal component loads.

Gray-box methods, which use pre-existing knowledge of the dynamical structure, have been used off-line with genetic algorithms (GA) [5] and recursively on-line (in real-time). Off-line methods may be good for initial model acquisition but on-line methods are desirable—buildings age and change physically over time due to deterioration or reconfiguration and change temporally as occupant usage evolves. On-line methods can also be dual-purposed as process monitoring, analysis, and fault detection devices.

Examination of traditional online gray-box techniques in buildings with adaptive control [2] and Extended Kalman Filters (EKF) [19] shows the need for further development. Historically, adaptive control techniques have fallen short because they require autonomous tuning of a complex forgetting factor, which requires actively monitoring the excitation level of incoming data as an information criterion used to enable or disable thermal model learning [2]. Many on-line methods [2], [4], and [5] have used single zone models for demonstrating utility, but in reality multi-zone simulations are necessary before any modeling method will gain wide acceptance because zone interactions affect BAS control systems [18].

In [11] we demonstrated the first published study of a scalable modeling and online estimation framework for multi-zone building states, parameters, and unmodeled dynamics. Since our study in 2012 several other researchers have validated our initial claims and highlighted new challenges.

Studies using real and simulated data demonstrated individual aspects of the proposed scalable modeling and estimation framework. Massoumy in 2013 [13] demonstrated the applicability of gray-box estimation with real data collected from Michigan Tech's new Lakeshore Center building–he used an off-line batch parameter estimation routine with online state estimates, validating the results of our EKF versus UKF comparison. Fux [12] demonstrated the relevance of 1R1C models to individual rooms or entire zones by generating accurate predictions from a simple single-zone model of an entire building using real data in an EKF. The study by Fux [12] also validated the concept of multi-mode learning and the importance of characterizing disturbances. Martincevic [14] used simulated data from IDA-ICE in a year-long study to demonstrate a 50-zone model that learned parameters without disturbances from a constrained UKF. The model had a prediction error of less than 1°C RMS error if no unmodeled disturbances were present.

Other researchers demonstrated the extensibility of the KF as an online estimation framework and presented important insight. Studies showed ground coupling was not necessary in RC models for certain scenarios [8], extended an EKF for fault diagnosis and monitoring with real-data from a supermarket [20], used an Ensemble KF to constrain parameters to physically realistic values [21], and showed the applicability of RC models as a design tool using offline parameter estimation [22]. Lin [23] showed the need for meaningful inter-zone excitation in order to guarantee satisfactory information content is present in measured data. Lin's result corresponded with our follow-up study using an UKF learned model for MPC [24]. Lin further noted that one simple plot demonstrating prediction accuracy versus measured data is meaningless in that it doesn't say anything about the model's robustness for control.

In summary, recent studies have shown RC models are applicable for thermal modeling of buildings and Kalman filtering can learn parameters for models of buildings using both simulated and real data. Our paper builds upon recent advancements with an explanation of unique parameterization and formalizes the multi-mode estimation technique with a generalized algorithm for learning any disturbance pattern. We conclude with a year-long study of simultaneous state, parameter, and disturbance estimation showing robust, meaningful prediction accuracy.

## II. PARAMETER ESTIMATION FORMULATION

### A. Thermal Model

A standard thermal network captures the dominant convection and conduction heat transfer modes and mass transfer occurring between zones inside and outside the building, while solar gain is treated as an unmodeled disturbance. Internal zone radiation is linearly approximated and lumped with convection and conduction [25]. The thermal network matches that commonly used in the community and provides a simple mechanism to explore simultaneous model and disturbance estimation. Subsequently developed filters use the thermal model but are not restricted to it. One could select a non-linear model incorporating HVAC dynamics and use it in the proposed estimation framework.

Convection, conduction, and mass transfer heat flux $q_i$ (watts) into zone $i$ is contributed from the temperature differential to connected adjacent node(s) $j$ divided by the thermal resistance $R_{ij}$ (degree/watt) plus an additive term $b_i$ (watts) representing disturbances. *(Note: unless otherwise mentioned, subscripts denote zones.)*

$$q_i = \sum_j (T_j - T_i)/R_{ij} + b_i$$

The heat flux $q_i$ and thermal capacity $C_i$ (joule/degree) affects the time-based temperature rate of change $\dot{T}_i$.

$$C_i \dot{T}_i = q_i$$

Substituting for $q_i$, the temperature rate of change of node $i$ due to connection(s) with node(s) $j$ and disturbance $b_i$ becomes

$$\dot{T}_i = \sum_j (T_i - T_j)/(R_{ij} C_i) + b_i/C_i. \tag{1}$$

The derived representation for temperature change due to heat transfer is mathematically analogous to voltage change due to current flow in a resistor-capacitor network. For visualization a simple 2-node example with two capacitances and one resistance is shown in Fig. 2.

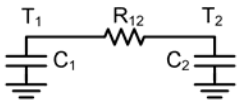

Fig. 2. Two node example thermal network.

Thermal radiation does play a significant role in the heating and cooling of many buildings. For the purposes of evaluating UKF parameter estimation and thermal load detection for a passive building, the radiation between surfaces and zones is linearly approximated and lumped with convection and conduction [25]. The proposed framework could readily be augmented for nonlinear radiation effects at the cost of increasing the number of associated parameters to learn.

Thermal disturbances significantly affect most buildings but are often overly complex to model requiring information about building geometry and neighboring foliage [26]. Solar gain is treated as an unmodeled external disturbance. This simplification removes complexities of modeling diffuse and direct sunlight, shading, and night sky radiation temperature, and allows for simple disturbance generation in EnergyPlus by turning on or off environmental radiation transfer. The solar gain provides us with a specific periodic disturbance to estimate with patterns. In practice this technique could estimate any number of disturbances if one has some information about the disturbance frequency, intensity, or timing such as dusk and dawn times or building occupancy times. Common examples amenable to disturbance pattern estimation include occupant body-heat, equipment, computers, electrical loads, lighting, and HVAC.

Using the 2-node example, a state space representation can be derived where $\bar{T}$ is a vector of temperatures; $A$ is a matrix of RC values; and $\bar{b}(t)$ is a vector of additive, independent, time-varying disturbances such as solar radiation.

$$\dot{\bar{T}}(t) = A\bar{T}(t) + \bar{b}(t)$$

$$A = \begin{bmatrix} -\dfrac{1}{R_{12}C_1} & \dfrac{1}{R_{12}C_1} \\ \dfrac{1}{R_{12}C_2} & -\dfrac{1}{R_{12}C_2} \end{bmatrix} \tag{2}$$

$$\bar{b}(t) = \begin{bmatrix} \dfrac{b_1(t)}{C_1} \\ \dfrac{b_2(t)}{C_2} \end{bmatrix}$$

Based on [27], an n-node thermal network can be formalized by defining a simple undirected weighted graph with: nodes $N \coloneqq \{1,2,\ldots,n\}$ assigned capacitances $C_i$ and temperatures $T_i$; edges $E \subset N \times N$ that connect adjacent nodes with weights $\{R_{ij} \forall (i,j) \in E$ such that $R_{ij} = R_{ji} \forall (i,j) \in E\}$ that are assigned resistances. For a general thermal network with $n$ nodes, the $A$ matrix is

$$A = \begin{cases} A_{ij} = 0 & \text{if } i \neq j, (i,j) \notin E \\ A_{ij} = \dfrac{1}{C_i R_{ij}} & \text{if } i \neq j, (i,j) \in E \\ A_{ij} = -\sum_{i \neq j} A_{ij} & \text{if } i = j \end{cases} \tag{3}$$

### B. Parameterization

A minimal set of independent parameters must be specified for filters to enforce the system dynamics during parameter estimation [28]. Over-parameterization causes unidentifiable parameter manifolds or extra degrees of freedom and can result in violation of dynamics constraints and physics laws such as conservation of energy. In machine learning and system identification, indeterminate degrees of freedom can cause overfitting where the model learns the noise instead of the dynamics of interest. In estimation theory, parameter observability requires that the Fisher information matrix is invertible—redundant parameters or over parameterization breaks this observability criterion resulting in an unobservable subspace [29].

Efficient and reliable parameter estimation requires estimating a minimal number of parameters [28]. From Equation (2) there are only two unique parameters required to describe the $A$ matrix despite it containing three variables—two resistances and one capacitance. The extra parameter acts as a scaling factor and can be quantified only if the heat flux $q$ is provided in addition to the temperature histories. Without the scaling factor only a time-constant can be inferred. Suppose the time-constant for our system was 1. Then $R_{ij}C_i = 1$ and we get the plot in Fig. 3 of possible values for $R_{ij}$ and $C_i$, many of which are violations of physics first principles such as a negative thermal capacitance and resistance. Unfortunately, removing the negative-valued parameter space does not resolve the ambiguity in selecting the true resistance and capacitance values for the provided time constant. This ambiguity generally makes the estimation problem numerically unstable, theoretically unobservable, or practically unreliable. Rectifying the ambiguity could be done with actual heat flux information which is generally unavailable in practice, so for this study, selecting a minimal set of parameters mitigates the problem.

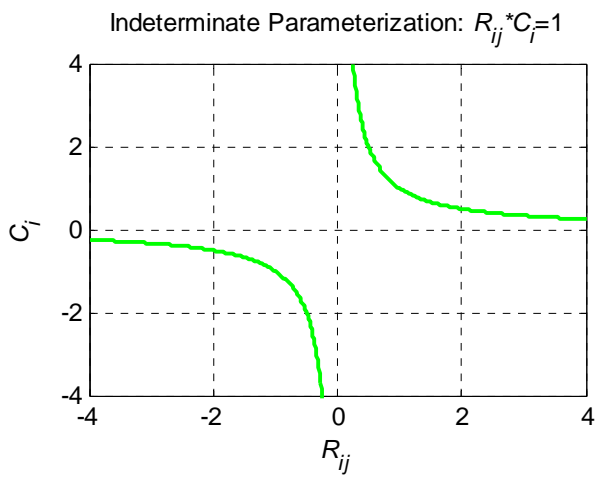

Fig. 3. Non-minimal parameterization gives rise to estimation ambiguity. The curve satisfying $R_{ij}C_i = 1$ demonstrates the unobservable subspace which includes physically meaningless negative quantities. $R_{ij}$ and $C_i$ cannot be uniquely identified.

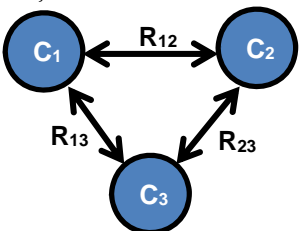

Fig. 4. Three-node graph with one loop.

Now we present methods based on a careful graph study to obtain a minimal parameter set for thermal network estimation. Because diagonal terms in $A$ are linear combinations of the off-diagonal terms, parameter estimation is only performed for off-diagonals. RC products are estimated together in order to reduce the non-linearity of the estimation problem. Parameterization of trees, graphs with no cycles of which Fig. 2 is an example, with combined RC products automatically guarantees a minimal representation of the system.

Unfortunately this minimal guarantee does not extend to graphs containing closed cycles. Fig. 4 is an example of a graph containing a cycle whose state space $A$ matrix is shown.

$$A = \begin{bmatrix} -\left(\frac{1}{R_{12}C_1} + \frac{1}{R_{13}C_1}\right) & \frac{1}{R_{12}C_1} & \frac{1}{R_{13}C_1} \\ \frac{1}{R_{12}C_2} & -\left(\frac{1}{R_{12}C_2} + \frac{1}{R_{23}C_2}\right) & \frac{1}{R_{23}C_2} \\ \frac{1}{R_{13}C_3} & \frac{1}{R_{23}C_3} & -\left(\frac{1}{R_{13}C_3} + \frac{1}{R_{23}C_3}\right) \end{bmatrix} \quad (4)$$

We arbitrarily selected $R_{13}C_1$ to show that one of the six RC products is redundant and can be eliminated by multiplying and dividing the other $R_{ij}C_i$ parameters by each other around the cycle:

$$R_{13}C_1 = \frac{R_{12}C_1 \times R_{23}C_2 \times R_{13}C_3}{R_{12}C_2 \times R_{23}C_3}$$

In a graph, each cycle which uses at least one unique edge and passes through no nodes with infinite capacitance may be used to eliminate one redundant RC product from the estimation problem by multiplying and dividing around the loop. For any thermal network the total number of unique parameters should be one less than the sum of the number of resistances and capacitances.

In general unique edges should be selected for elimination. Eliminating a shared edge between two cycles joins the two cycles mathematically through multiplication in the estimation routine which can negatively impact numerical stability. Selecting multiple redundant parameters to prune from a graph estimation problem should be done such that each redundant RC parameter lies on a globally unique edge for its respective cycle, and the shortest available cycle should be chosen for calculation in order to guarantee minimal parameter cross-sensitivity.

Two nodes that have no shared conduction or convection are considered independent, and any edge directly connecting them is pruned from the graph to give the simplest representation. Independent ambient nodes such as external temperatures have infinite capacitance in the thermal network. External nodes may have unique update functions depending on the simulation and weather desired for the modeling exercise. As an example look back at Fig. 2, if $T_2$ were an external temperature, setting $C_2 = \infty$ would give the following state space representation.

$$A = \begin{bmatrix} -\frac{1}{R_{12}C_1} & \frac{1}{R_{12}C_1} \\ 0 & 0 \end{bmatrix} \quad (5)$$

$$\bar{T} = [T \quad T_{ext}]^T$$

## C. Extended Kalman Filter (EKF)

For all tests the state space system is integrated at one minute interval time-steps with Euler integration to allow discrete-time filter implementation. Parameter estimation with the Kalman Filter is achieved by augmenting the temperature states $\bar{T} = [T_1 ... T_n]^T$ with unique parameters $\bar{p} = [(1/RC)_1 ... (1/RC)_k]^T$ and disturbances $\bar{b} = [b_1/C_1 ... b_l/C_l]^T$ together in the state representation $\hat{x} = [\bar{T}^T, \bar{p}^T, \bar{b}^T]^T$. For the purposes of estimation, the full discrete-time stochastic system is

$$\begin{aligned}
\bar{T}(k+1) &= A(p(k))\bar{T}(k) + \bar{b}(k) + \overline{w_1}(k) \\
\bar{p}(k+1) &= \bar{p}(k) + \overline{w_2}(k) \\
\bar{b}(k+1) &= \bar{b}(k) + \overline{w_3}(k) \\
\bar{z}(k) &= \bar{T}(k) + \bar{v}(k)
\end{aligned} \quad (6)$$

where $\overline{w_1}(k)$ represents process noise, $\overline{w_2}(k)$ represents estimation uncertainty in RC parameters, $\overline{w_3}(k)$ represents process noise for disturbances, and $\bar{v}(k)$ represents measurement noise. All noise terms are assumed zero mean, Gaussian, white, and stationary. Note that in (6), the matrix $A$ is actually a vector of functions composed from $\bar{p}(k)$ as defined by the parameterization of the RC terms. This representation results in multiplication and division of estimated parameters through the dynamics function. Specifically, temperature is being multiplied by RC parameters necessitating non-linear estimation techniques.

For baseline comparisons, an EKF and UKF are formulated. In order to define notation, the prediction and update steps of the discrete EKF are shown in (7), but for proper treatment of the derivation and background of the Kalman Filter, please consult [29], [30], [31]. *(Note: For brevity in the following Kalman Filter formulations, notation deviates from the modeling section: subscripts denote time rather than node indices.)*

*Predict:*
$$\begin{aligned}
\hat{x}_{k|k-1} &= f(\hat{x}_{k-1|k-1}, u_k) && \text{State Estimate} \\
P_{k|k-1} &= F_k P_{k-1|k-1} F_k^T + Q_k && \text{State Covariance}
\end{aligned}$$

*Update:*
$$\begin{aligned}
\tilde{y}_k &= z_k - h(\hat{x}_{k|k-1}) && \text{Innovation} \\
S_k &= H_k P_{k|k-1} H_k^T + R_k && \text{Innov. Covariance} \\
K_k &= P_{k|k-1} H_k^T S_k^{-1} && \text{Optimal Kalman Gain} \quad (7)\\
\hat{x}_{k|k} &= \hat{x}_{k|k-1} + K_k \tilde{y}_k && \text{State Estimate} \\
P_{k|k} &= (I - K_k H_k) P_{k|k-1} && \text{State Covariance}
\end{aligned}$$

*Jacobians:*
$$F_k = \left.\frac{\partial f}{\partial x}\right|_{\hat{x}_{k-1|k-1}, u_{k-1}}, \quad H_k = \left.\frac{\partial h}{\partial x}\right|_{\hat{x}_{k|k-1}}$$

For the filter dynamics function $f()$, temperature state dynamics follow the previously derived thermal model while the RC and disturbance parameters are modeled as constants. Artificial process noise for the constant parameters, denoted $\overline{w_2}(k)$, allows the filter to change its estimate of these values through time and allows the filter to track the true time varying disturbance. The set of process noise terms $\overline{w_1}(k)$, $\overline{w_2}(k)$, and $\overline{w_3}(k)$ are stacked as defined by the state $\hat{x}$ and drawn from distribution $Q$. The measurement noise terms $\bar{v}(k)$ are drawn from distribution $R$.

The measurement function of the thermal network is linear so $H$ is simply an *n* row identity matrix (provided all temperatures are measured) padded with columns of zeros for the parameters. The update portion of the filter therefore can be written as a simple linear Kalman Filter. However in the prediction step, calculation of the Jacobian matrix $F$ depends whether one chooses to estimate *RC* or *1/RC* parameters. A comparison of both cases demonstrates that *RC* parameters would be a poor choice, especially during the initial acquisition stage, because poor parameter estimates will be squared and could easily cause the filter to diverge and blowup.

$$\begin{aligned}
\text{Case 1:} \quad & p = RC \\
& \frac{\partial f_i}{\partial p_j} = \frac{\partial}{\partial p_j}\left(\frac{1}{p_j} T_k\right) = -\frac{1}{p_j^2} T_k \\
\text{Case 2:} \quad & p = \frac{1}{RC} \\
& \frac{\partial f_i}{\partial p_j} = \frac{\partial}{\partial p_j}(p_j T_k) = T_k
\end{aligned} \quad (8)$$

Thus, for numerical stability the EKF must follow Case 2 and estimate *1/RC* parameters.

Measurement noise is specified based on the accuracy of the temperature sensors. Process noise is specified for the temperature states based on the level of zone aggregation used while the *RC* and disturbance process noise is set to an artificial value greater than zero in order to allow the filter to vary its estimate of these parameters through time. Increasing process noise level for any parameter indicates that the model isn't confident of its ability to describe the process evolution of that parameter. Disturbances, which by their nature the model is not explicitly capturing, are biased and vary with time. In order to estimate the disturbances over time; their noise level is set to be non-zero. Because the RC values should be fairly constant while the disturbance bias may change throughout the course of a day, the noise level for RC parameters should be much smaller than disturbances.

## D. Unscented Kalman Filter (UKF)

Unlike the EKF, which uses a Jacobian first order linearization evaluated at the current estimate to propagate a probability distribution through a non-linear transform, the UKF uses the Unscented Transform to pass a distribution through a nonlinear transform. Specifically the UKF samples *(2n+1)* points in the distribution, evaluates each point through the non-linear transform and then recombines these points to generate a transformed mean and covariance which is oftentimes more accurate and stable than that obtained from the single point EKF linearizations [32]. The samples, called sigma points, are evenly spaced to capture at least the first and second order moments of the distribution and are weighted such that the covariance and mean of the samples matches that of the original distribution. After being mapped through the non-linear transform the resulting points are multiplied by their assigned weights to determine the transformed mean and covariance. For linear systems both the EKF and UKF perform identically to a traditional Kalman Filter but for certain non-linear systems a UKF can provide higher accuracy with the same order of calculation complexity.

The UKF was implemented with the same augmented state vector and used the same measurement and process noise values as the EKF. Removing the requirement to design and calculate a Jacobian, the UKF is amenable to either *RC* or *1/RC* parameter representations—both are evaluated in the results section. The standard values of $\alpha = 10^{-3}$, $\kappa = 0$, $\beta = 2$, typical for a Gaussian distribution, were used to generate the following samples $\chi$ and weights $W$.

$$\lambda = \alpha^2(L + \kappa) - L$$
$$\chi_{0,k-1|k-1} = \hat{x}_{k-1|k-1}$$
$$\text{For: } i = 1, \ldots, L$$
$$\chi_{i,k-1|k-1} = \hat{x}_{k-1|k-1} + \left(\sqrt{(L+\lambda)P_{k-1|k-1}}\right)_i$$
$$\text{For: } i = L+1, \ldots, 2L$$
$$\chi_{i,k-1|k-1} = \hat{x}_{k-1|k-1} - \left(\sqrt{(L+\lambda)P_{k-1|k-1}}\right)_{i-L}$$
$$W_0^{(m)} = \frac{\lambda}{L+\lambda}$$
$$W_0^{(c)} = \frac{\lambda}{L+\lambda} + (1 - \alpha^2 + \beta)$$
$$\text{For: } i = 1, \ldots, 2L$$
$$W_i^{(m)} = W_i^{(c)} = \frac{1}{2*(L+\lambda)}$$

The samples were then recombined to give the *a priori* state and covariance estimates.

$$\chi^{k|k-1} = f(\chi^{k-1|k-1})$$
$$\hat{x}_{k|k-1} = \sum_{i=0}^{2L} W_i^{(m)} \chi_i^{k|k-1}$$
$$P_{k|k-1} = \sum_{i=0}^{2L} W_i^{(c)} [\chi_i^{k|k-1} - \hat{x}_{k|k-1}][\chi_i^{k|k-1} - \hat{x}_{k|k-1}]^T$$

Note that $\lambda$ and $W$ can be reused so only the $\chi$ terms need to be recalculated each iteration. Because the measurement function is linear, the unscented transform is only used in the prediction step; the measurement step is simply calculated as a linear Kalman filter update.

### E. EKF vs. UKF

In order to compare an EKF and UKF the 2-node thermal network from Fig. 2 was chosen. Looking back to (5), the system can be parameterized as $A_p$.

$$A_p = \begin{bmatrix} -p_1 & p_1 \\ 0 & 0 \end{bmatrix}$$
$$\bar{T} = [T \quad T_{ext}]^T$$

In order to compare both the parameter estimation and bias detection capabilities only $T$ was sensed—$T_{ext}$ was treated as a constant latent variable. A second parameter $p_2$ was used to estimate $T_{ext}$ as a linear additive disturbance as shown.

$$\dot{T} = -p_1 T + p_1 T_{ext}$$
$$p_2 = p_1 T_{ext} \quad (9)$$
$$\dot{T} = -p_1 T + p_2$$

A 10,000 run Monte Carlo simulation was conducted to compare the EKF and UKF performance for the simple two-parameter search problem. Table 1 shows the distributions for all parameter and filter values which were randomly sampled at the beginning of each test run. The distributions of parameter and temperature values were chosen such that some runs will have high levels of excitation while others will have no excitation—effectively $T(0)$ equal to $T_{ext}$. Results of the simulation, shown in Table 2, demonstrated that the UKF outperformed the EKF for non-linear estimation of parameter $p_1$ but had statistically similar performance for estimation of linear parameter $p_2$. The UKF showed filter stability—resilience to exponential tracking divergence—while 2% of the EKF runs went unstable and did not complete execution. These results agree with published studies comparing the filters' general performance in other application areas [33], [34]. The EKF could estimate linear additive disturbances, which is in agreement with [19], but when estimating coefficient parameters such as weights in a thermal network, the UKF is a more viable solution.

Further testing showed that increasing the number of temperature zones or trying to directly estimate *RC* instead of its reciprocal, *1/RC*, further degraded the EKF stability and performance. However for the UKF, comparison of estimating *RC* products and their reciprocals showed that direct *RC* parameters are more robust. For zones with little thermal connection, the filter estimating *RC* parameters will continue increasing estimates but will be finitely decreasing the covariance, so the estimate will eventually converge.

When *1/RC* parameters are utilized for the same zones, the estimate will be driven to zero, but the covariance will not

TABLE I
MONTE CARLO SAMPLED PARAMETER DISTRIBUTIONS

| Variable | Quantity | Range |
|---|---|---|
| $p_1$ | Truth | $\mathcal{U}(0,1)$ |
| $p_2$ | Truth | $\mathcal{U}(0,20)$ |
| $\hat{p}_1(0)$ | I.C. Estimation | $\mathcal{U}(0,1)$ |
| $\hat{p}_2(0)$ | I.C. Estimation | $\mathcal{U}(0,20)$ |
| $T(0)$ | Truth | $\mathcal{U}(0,100)$ |
| $\hat{T}(0)$ | I.C. Estimation | $T(0) + \mathcal{N}(0,10^{-3})$ |
| $R$ | True Meas. Error | $|\mathcal{N}(.5,.5)|$ |
| $\hat{R}$ | Est. Meas. Error | $|\mathcal{N}(.5,.5)|$ |
| $Q$ | True Process Variance | $diag\{1,0,0\}$ |
| $\hat{Q}$ | Est. Process Variance | $diag\begin{Bmatrix}1, \mathcal{U}(0,10^{-3}), \ldots \\ \mathcal{U}(0,10^{-3})\end{Bmatrix}$ |
| $P(0)$ | Initial Est. Variance | $diag\begin{Bmatrix}1, \mathcal{U}(0,5\times 10^{-3}), \ldots \\ \mathcal{U}(0,5\times 10^{-3})\end{Bmatrix}$ |

$\mathcal{U}$ = Uniform Distribution, $\mathcal{N}$ = Normal Distribution.

TABLE 2
10,000 RUN MONTE CARLO RESULTS

| Test | Statistic | Result |
|---|---|---|
| EKF | Numerical Instability Count | 175 |
| EKF | $p_1$ Average Estimation Error | 3.16 |
| EKF | $p_2$ Average Estimation Error | 60.5 |
| UKF | Numerical Instability Count | 0 |
| UKF | $p_1$ Average Estimation Error | 0.18 |
| UKF | $p_2$ Average Estimation Error | 58.0 |

decrease as quickly, which can result in a zero crossing that violates conservation of energy by estimating a negative parameter. As a result, *RC* parameters were estimated by the UKF for all remaining tests.

*F. Disturbance Estimation and Pattern Recognition*

Direct sensing of disturbance heat flux is rarely practical in building systems. However, timing information for disturbances is typically available, so a practical method is presented to learn disturbances given only timing information with no prior disturbance quantification. This method could readily be augmented if additional disturbance heat flux data was available.

Looking at thermal model (1), a change in zone temperature may be explained away using either connected zones with their respective temperatures or additive disturbances. In order to get satisfactory estimation performance this ambiguity must be considered in the filter design lest it manifest problems akin to non-minimal parameterization. The engineering solution selected to rectify the problem, splits the estimation problem based on the presence of disturbances and manipulates the process covariance for estimated disturbance parameters based on timing of expected unquantified disturbances. From a control theory perspective the system does not have time-invariant observability. However, buildings are a time-varying system that have some periodicity. Looking over a horizon, for example one day for solar disturbances, we can learn constant parameters when no disturbances are present and learn disturbances after having estimated the constant parameters. This partitioning enables time-varying observability.

The presented method is not claimed to be optimal, rather it is a practical solution based on engineering judgment of typical scenarios common in buildings. Typically, a system can sense if people are using a building but cannot measure their heat flux or the equipment they use, likewise it can sense if the HVAC is on but not the exact heat flow delivered to a specific room. The approach attempts to use commonly available timing knowledge to quantify and infer disturbances that are not directly measurable and only partially predictable.

Learning of disturbances was done in a Markov fashion: the estimator assumed no knowledge of previous historical disturbance patterns and estimated a new disturbance value $b_i$ at each time step based on the previous time step's estimate, the dynamic model, and the current measurement. The disturbance states in the UKF were modeled as constants which have zero-mean Gaussian additive noise. A characteristic change in the disturbance such as a heater turning on or the sun coming up at dawn violates the zero-mean assumption causing a bias in the disturbance. In order to track these sudden bias changes using a simple UKF, the variance(s) correlating to those specific zone(s) disturbances were inflated to allow the filter to acquire and track the new value. This artificial tuning of the covariance is similar to tuning a forgetting factor in adaptive control frameworks.

In the Matlab-based simulation, heating or cooling was arbitrarily added to individual zones from 10am to noon, so the covariance was increased around those times. In the EnergyPlus simulation, the primary unmodeled disturbance was solar radiation, so the covariance was increased around dusk and dawn. This variance tuning is visually depicted in Fig. 5.

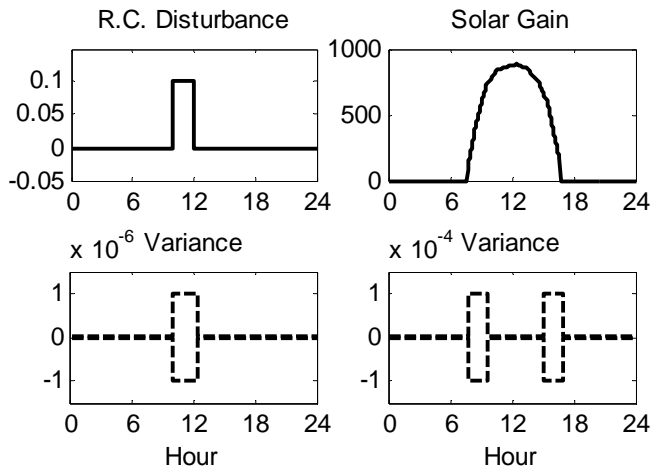

Fig. 5 Disturbance Parameter Process Variance Tuning

Because the UKF can explain temperature swings by either tuning RC or disturbance values, a multi-mode approach was chosen as a uniform method to split the estimation problem. In order to acquire good estimates, the UKF was operated in two modes: A) Acquisition Mode: initially only RC products were estimated by running the filter at night when solar gains were at a minimum, and B) Monitoring Mode: both RC products and disturbances were estimated simultaneously after RC product estimates had started to converge to constant values. This splitting proved critical to obtaining good estimates from EnergyPlus data but unnecessary for the simple Matlab based simulation due to its consistently high level of external temperature excitation.

Disturbance estimates from the entire multi-day learning period were heuristically combined in order to generate a 24-hour pattern. This disturbance pattern was then used when predicting the building's thermal response.

The heuristic pattern recognition algorithm was a simple weighted average. For thermal network simulated data the true disturbances were identical every day. Thus the bias values learned at each time step were equally averaged to generate one 24-hour pattern for each bias term. Mathematically this can be written as summation over $n$ days of learning with minute time steps to generate a disturbance pattern $d_i$ correlating to zone $i$ based on the UKF estimated bias $b_i$ as shown, where the value inside brackets notates the minute-based time step index.

$$d_i[k] = \sum_{j=1}^{n} b_i[j,k] \; \forall k \in [1 \; to \; 24 \times 60] \qquad (10)$$

The predominant unmodeled disturbance from EnergyPlus data was radiation from the sun, which is dependent on the cloud cover, time of day, season of year and other factors. For the purpose of engineering a robust simple solution we made a realistic assumption that we had a measurement of the average solar intensity for morning and afternoon and used this for both pattern recognition and simulation predictions. The solar intensity reading was calculated as the summation of the Environment Direct Solar and Environment Diffuse Solar variables from EnergyPlus. Algorithm 1 contains the weighted average and prediction steps used for EnergyPlus simulations.

ALGORITHM 1

♦ INITIALIZATION
-Max solar radiation: $r_{max}$
-Indices: zone $i$, day $j$, time of day $k$
-Dawn, midday, and dusk times: $dawn[j]$, $midday[j]$, $dusk[j]$
-Morning & afternoon avg solar radiation: $morning[j]$, $afternoon[j]$
-Estimated bias per zone $i$: $b_i[j,k]$
-Measured solar radiation intensity: $r[j,k]$
♦ WEIGHTED DISTURBANCE PATTERN
-Disturbance Pattern: $d_i[k]$
for $j = 1$ to $n$
  if $(morning[j] + afternoon[j])/2 > 0.35 \times r_{max}$ then
    for $k = 1$ to $24 \times 60$
      if $k < dawn[j]$ then $d_i[k] \mathrel{+}= b_i[j,k]$
      elseif $k < midday[j]$ then $d_i[k] \mathrel{+}= b_i[j,k]/morning[j]$
      elseif $k < dusk[j]$ then $d_i[k] \mathrel{+}= b_i[j,k]/afternoon[j]$
      else $d_i[k] \mathrel{+}= b_i[j,k]$
    end
  end
end
$d_i[k] = d_i[k]/n$
♦ PREDICTED DISTURBANCE
-Bias for Prediction: $d_i[k]$
for $j = n + 1$ to ...
  for $k = 1$ to $24 \times 60$
    if $k < dawn[j]$ then $b_i[k] = d_i[j,k]$
    elseif $k < midday[j]$ then $b_i[k] = d_i[j,k] \times morning[j]$
    elseif $k < dusk[j]$ then $b_i[k] = d_i[j,k] \times afternoon[j]$
    else $b_i[k] = d_i[j,k]$
  end
end

By averaging the solar intensity before midday and after midday, two weights were determined for each day. In order to ensure sufficient signal to noise ratio for disturbance pattern estimation, a day's disturbance estimates were discarded if the day's total solar intensity averaged below 35% of the maximum solar intensity possible for that location. The remaining days had their bias estimation values multiplied by the ratio of maximum solar radiation to respective half day average measured intensity in order to normalize the bias values. Then the normalized bias values were averaged using the same equation for $d_i$ to estimate the repeated daily disturbance profile. When used for predictions, the disturbance profile was scaled by the predicted solar intensity weights to predict each zone's unique solar disturbance quantity at each timestep throughout the day. Thus predictions utilized the final estimated RC parameter, 24-hr disturbance patterns, predicted external temperature profile, and half-day average predicted solar intensity. The pattern recognition shown in Algorithm 1 could be applied to any cyclic disturbance. The variables $k$, $midday$, $dusk$, and $morning$ would be adapted to whatever period and timing knowledge a designer had of other loads or disturbances present in the environment, and the arrays $dawn$ and $afternoon$ would be modified to contain the pattern array.

## III. SIMULATION RESULTS

### A. UKF 5-room Simulated Performance

A six-node thermal network, corresponding to five internal zones and one external temperature shown in Fig. 6, was used to evaluate the UKF parameter estimation and thermal disturbance detection. For this first evaluation, measurement data was generated from a model whose dynamics were structurally identical to the dynamics used in the UKF. The 5-room models shown here are based on that in [11], but feature extended explanations, derivations, and simulation results.

Given five finite capacitances and thirteen resistances, there were a total of seventeen unique *RC* products for the UKF to estimate in this model. A first test was run in acquisition mode, so disturbance estimation was disabled. The external temperature forcing function was composed from the sum of a 40 degree peak-to-peak sinusoid with period of one day, a 10 degree peak-to-peak sinusoid with period of 4 hours, and random noise which would allow the temperature to drift from day to day. Resistance and capacitance values were chosen such that thermal lag in the simulation would be similar order as thermal lag in a small to medium sized building. Temperature states were initialized with less than a degree of error while *RC* estimates were all arbitrarily initialized to 1000 with a standard deviation of 500. Using four days of recorded data, *RC* parameters were learned by the UKF. At the end of the four day simulation, a 48 hour prediction of the five zones' temperatures was made using the acquired *RC* parameters. The true 48 hour external temperature profile was provided to both the true dynamics and UKF dynamics in order to establish a fair comparison baseline for evaluation of the UKF. In a real system inaccuracies in the weather forecast would degrade the prediction quality.

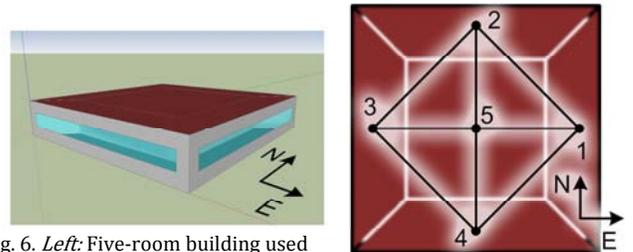

Fig. 6. *Left:* Five-room building used for simulations, *Right:* Top view of node labeled internal thermal network representation. White lines represent building internal walls. A sixth unlabeled node acts as an external temperature

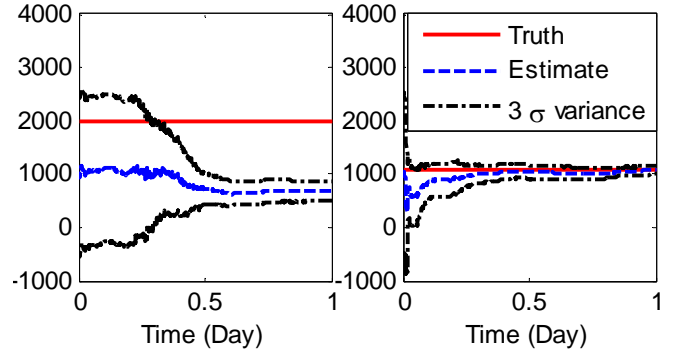

Fig. 7. Example estimation of two parameters from 5-room building using RC-thermal model data.

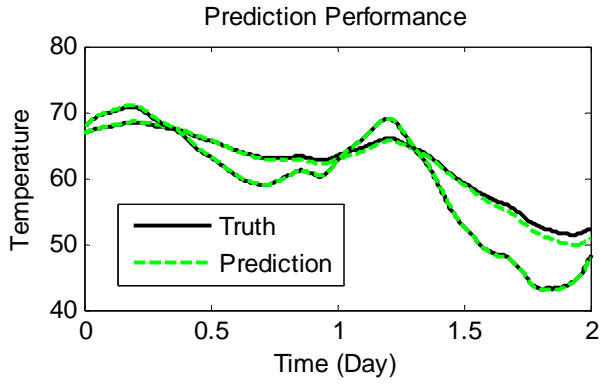

Fig. 8. Example 48-hour prediction window in a disturbance free environment based on RC estimates. External temperature forcing function was composed from two sinusoids plus white noise. (Deg F)

Running the described simulation showed that some parameters are estimated very well while others are not; two characteristic examples of this are shown in Fig. 7. The external temperature does not fully excite all of the node-to-node thermal connections which limits the filter's ability to precisely determine all of the parameters. Despite this numerical estimation error, the 48-hour prediction demonstrated excellent matching between the UKF estimated model and the true model. From a poorly performing filter one would expect predictions to have increasing divergence or lag as the prediction window is increased. Characteristic temperature predictions of the highest and lowest capacitance rooms are shown in Fig. 8 for visualization. In our chosen model the poorly estimated parameters correlate with higher order dynamics that need-not be accurately estimated for good model fitting. For practical applications this is analogous to having multiple zones that are always excited together such that their relative interaction need not be known for useful predictions.

For a second evaluation daily repeating disturbances were introduced uniquely into each zone of the house. The disturbance states in the UKF were modeled as constants which have zero-mean Gaussian additive noise; in order to track a sudden bias change using a simple UKF, the variance correlating to that specific bias must be inflated to allow the filter to acquire and track the new value. This inflation must occur anytime the disturbance changes significantly enough that the zero mean Gaussian assumption is severely violated, such as when an HVAC system turns on or at dusk and dawn due to solar radiation. For the combined *RC* and disturbance estimation, 4 days of data were again used for training. Disturbances turned on at 10:00 and off at 12:00 each day, so the estimate covariance *P* was boosted at those times and the process variance was also temporarily increased when the disturbance was cycled on and off as previously shown in Fig. 5. With this variance tuning method, the filter had excellent disturbance tracking and similar *RC* estimation accuracy. Example tracking of two disturbances is shown in Fig. 9. Using the described simple heuristic pattern recognition with a 24 hour period shown in Equation (10), the estimated biases from all four days were averaged to generate a single day estimate of the repeating disturbance pattern. A 48-hour prediction was then made using the average disturbance and final *RC* estimates along with the exact external temperature profile. Again, excellent predictions were made with the estimated model. Fig. 10 plots temperatures of two zones comparing predictions from the RC only estimation model and the RC plus disturbance estimation model to the truth model. This evaluation provides good indication of the applicability of the UKF to thermal network parameter and disturbance estimation.

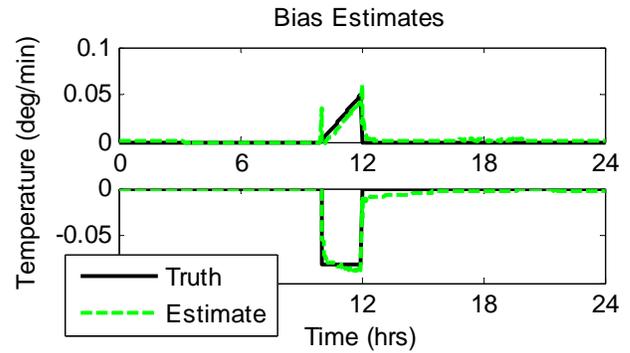

Fig. 9. The actual disturbances introduced into system for the two zones shown in Fig. 10 are labeled above as "Truth" and were repeated on a 24 hour cycle. Dashed lines represent the UKF 4-day average estimate of the disturbances which was used for predictions in Fig. 10. (Deg F)

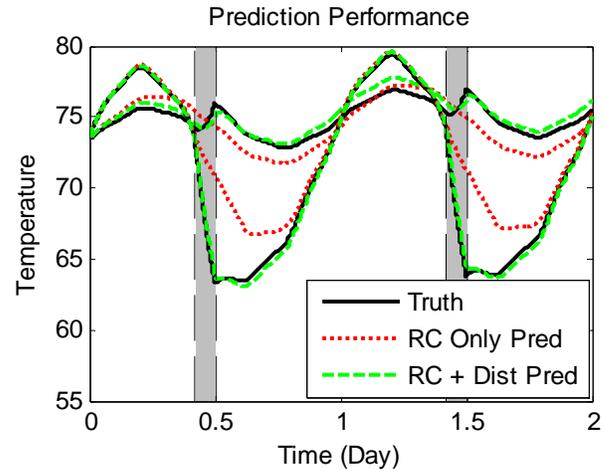

Fig. 10. Example 48-hour prediction derived from RC estimates and estimated daily cyclic disturbances. Gray boxes denote times where external disturbances were present in the true system. An external temperature forcing function was composed from two sinusoids plus low amplitude white noise. (Deg F)

### B. UKF EnergyPlus Performance

Given the excellent performance of the UKF on data generated by the thermal network model, the UKF was tested on data generated from an EnergyPlus simulation. Individual data traces show typical estimation performance which is validated in an aggregated year-long demonstration. This more realistic EnergyPlus model, shown in Fig. 6, had five rooms correlating to the five zones; realistic data for the floor, wall, and ceiling composition; and windows on the four exterior walls, which pointed in the cardinal compass directions. The structure was simulated with weather and solar radiation for Elmira, NY. In order to acquire good estimates, the UKF was operated in two modes: A) Acquisition Mode: night-only estimation of RC products, and B) Monitoring Mode: simultaneous estimation of both RC products and disturbances. Distinguishing learning modes was less of a concern for the thermal network simulated data because the disturbances were only on for short periods of time and the external temperature had a consistently high level of variation to excite the system. Because the primary unmodeled thermal disturbance was solar radiation, covariance tuning was done for the Monitoring Mode by increasing the bias states' process variance for two hours starting at dawn and another two hours ending 20 minutes after dusk as previously shown in Fig. 5.

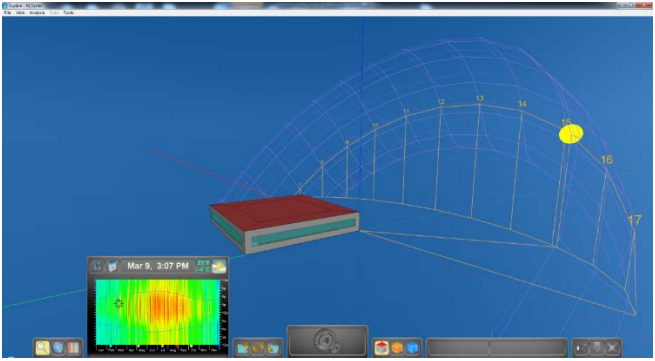

Fig. 11 Screenshot from Sustain showing afternoon sun on West Wall

Fig. 11 shows the path of the sun on March 9th and the variability of the sun path over the course of the year that will be simulated by EnergyPlus. The graphic was generated by Sustain, a front-end for EnergyPlus developed by researchers at Cornell University Program of Computer Graphics [35]. Due to the axis-inclination of the Earth, the sun's path and solar gain varies over the year. Fig. 13 shows the resulting average disturbances from a four day test where biases were only estimated for the last two days and then combined into a 24-hour pattern. Notice how the East room is heated in the morning, the West room is heated in the afternoon, and the South room is heated all day, which correlates nicely with expected heat from solar radiation. Plots of the 48-hour predictions are shown in Fig. 12. Predictions which utilize solar disturbances have much higher accuracy than the *RC* only predictions. The accuracy of these predictions ground assumptions made in the thermal network formulation and more importantly demonstrate the utility of the UKF for system identification of buildings' thermal envelope.

### C. UKF EnergyPlus Year Study

Further analysis of the EnergyPlus generated data was conducted by analyzing a total year of data. A unique UKF instance was initialized each day on the first 357 days of the year and run for 7 days, 3 days in acquisition mode and 4 days in monitoring mode. Then the UKF was used to predict the building's response on the eighth day and compared against the buildings actual response from EnergyPlus. This generated 357 sets of learned parameters, bias estimates, and 24-hour prediction simulations.

Of the total set, 43 simulations resulted in estimation routine errors such as negative parameter estimates or covariance shrinking to zero causing an UKF matrix inverse calculation failure. From further analysis, simple estimation monitoring by a human or addition of heuristic rules to the existing framework would fix all 43 estimation routine failures. For example over 10 of the failures occurred because 4 consecutive days had less than 35% solar radiation causing no disturbance pattern to be learned. Fixing these sorts of numerical issues to guarantee 100% reliability in an automated algorithm is outside the scope of the current study. Results suggested the algorithm could easily be matured for practical application by adding a number of heuristics.

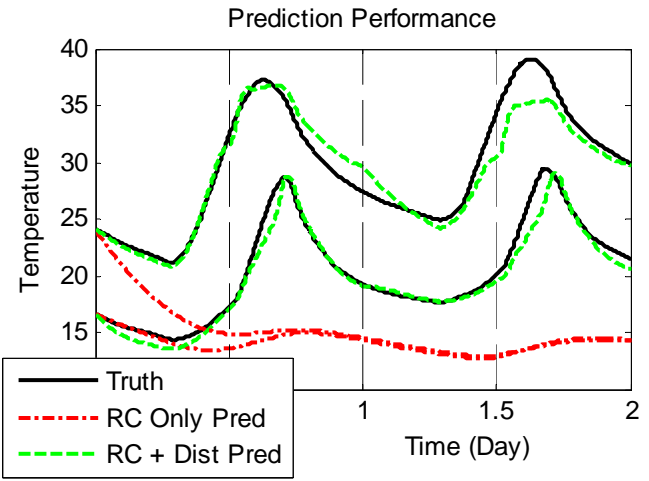

Fig. 12. *(Top)* South Zone, *(Bottom)* West Zone of 48-hour EnergyPlus predictions (Deg C)

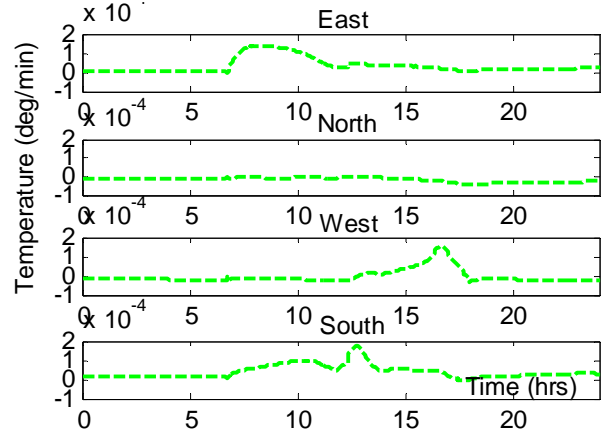

Fig. 13. Estimated 24-hour cyclic disturbances from EnergyPlus data (Deg C).

Using the 314 successful estimation runs, we compared the 24-hour prediction simulations against the buildings' truth simulation from EnergyPlus and found good accuracy—the models have enough fidelity to be used for control. Over a 12 hour prediction horizon the root mean square (RMS) temperature prediction error was 1.16°C and over 24 hours the RMS temperature prediction error was 1.48°C. ASHRAE standards mandates that vertical temperature stratification in an occupied zone should be less than 5.4°F (3°C) [36]. Home and office thermostats often use a dead-band of 4°F (2.2°C) to 8°F (4.4°C). The model's prediction errors are well within these design bounds for the 24 hour prediction horizon. In Fig. 14 the RMS error for the prediction is shown over time demonstrating good performance. Increased learning periods of 14 and 28 days further reduced prediction error but did not drive it to zero—likely because the simple RC model could not capture the entire fidelity of the EnergyPlus truth simulation.

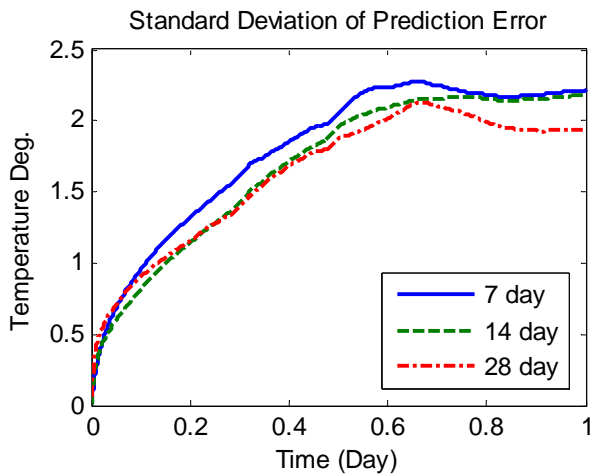

Fig. 14. Root Mean Square error of temperature predictions of all 5 zones over the year for different lengths of estimation learning time. (Deg C)

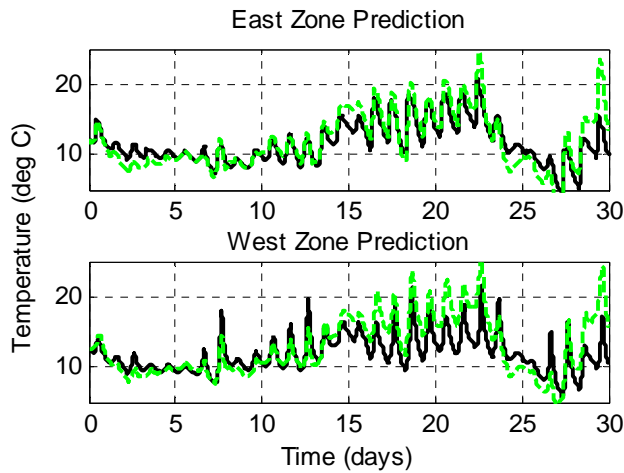

Fig. 15 Month prediction using a model learned from 1 week of data.

Additionally Fig. 15 shows a month-long stable prediction of the East and West zones based on a model learned from 7 days of data. This long prediction used the same data types as that in Fig. 12: correct zone initial temperature conditions, correct external temperatures over the horizon and half day average solar intensity values over the horizon. The long-horizon prediction accuracy demonstrates the learned model is unbiased, stable, and robust. To the author's knowledge, this is the first year-long study of an online UKF estimating disturbances with parameters and states for a building.

## IV. DISCUSSION

Results of the UKF estimation and model prediction capabilities have demonstrated the method as a powerful tool for thermal modeling of building systems. The simplicity with which a thermal network can be described combined with the numerical stability and robustness of the UKF are important factors which could enable its deployment as a scalable system identification routine for buildings thermal envelopes.

No physics constraints were applied to ensure RC parameters were positive, or to inform the bias and disturbance estimation. Realistic estimation of values was solely dependent on the quality of the chosen thermal model representation, estimation technique, and measured data. The authors expect that good results obtained from this paper's simulations would reflect realistic expectations of good perfomance in real world applications. Accurate bias tracking was achieved though covariance tuning, but this might not be scalable to certain buildings where disturbances occur on erratic schedules, so multiple hypothesis estimation or constraints may be augmented with a UKF to provide a more powerful solution.

Further investigation into model selection and fidelity could lead to performance improvements for the UKF depending on the target application and available computation and sensing hardware. For example, Dobbs [37] compared accuracy of thermal models across different levels of RC zone aggregation—leveraging such a tool may aid control-oriented model creation for the UKF. Extensions to the UKF may offer new opportunities for fault detection and monitoring [38].

One outstanding challenge remaining with this online estimation technique is a demonstration of the learned models' performance with model predictive controllers in practice. Some studies [13], [24], [23] have begun investigating how the quality of the learned model affects the performance of predictive controllers that use the model. The consensus to date is that intra-zone excitation is necessary in order to learn a building's internal coupling. Before controlling the building in a novel way to maximize energy savings, a buildings internal thermal coupling must be known. Deriving methods to monitor the quality of the measured data and better learn the building's thermal dynamics on-demand, by experimentally exciting the building, are the subject of a future paper by the authors.

## V. CONCLUSION

A multi-mode implementation of a multi-zone UKF was presented as a scalable and rapidly deployable system identification routine for building thermal dynamics. Using a passive 5-room model, the UKF demonstrated the ability to learn both dynamics parameters for a thermal network and unknown disturbances. 24-Hour predictions from UKF estimated parameters yielded accurate results which were validated with EnergyPlus simulations using a full year of data. The UKF, a data-driven, model-based approach, amenable to augmentation with numerical methods, provides a promising step towards a scalable framework to realize advanced BAS predictive controllers.

ACKNOWLEDGMENT

We would like to thank Dave Bosworth for his help generating an EnergyPlus 5-room building dataset for evaluation of parameter estimation methods.